\documentclass[runningheads]{llncs}
\usepackage[T1]{fontenc}
\usepackage{graphicx,verbatim}
\usepackage{amsmath}
\usepackage{booktabs}
\usepackage{tabularx} 
\usepackage[colorlinks=true, linkcolor=blue, urlcolor=blue, citecolor=blue]{hyperref}

\newcolumntype{Y}{>{\centering\arraybackslash}X}
\begin{document}
\title{FINDER: Zero-Shot Field-Integrated Network for Distortion-free EPI Reconstruction in Diffusion MRI}
\titlerunning{FINDER: Zero-Shot Distortion-free EPI Reconstruction}
%
\author{Namgyu Han\inst{1,\dagger}\orcidID{0009-0003-3292-7409} \and
Seong Dae Yun\inst{2,\dagger}\orcidID{0000-0001-7398-1899} \and
Chaeeun Lim\inst{1}\orcidID{0009-0005-4858-2155} \and
Sunghyun Seok\inst{1}\orcidID{0009-0004-8478-2528} \and
Sunju Kim\inst{1}\orcidID{0009-0007-0036-9405} \and
Yoonhwan Kim\inst{1}\orcidID{0009-0001-6770-2959} \and
Yohan Jun\inst{3,4}\orcidID{0000-0003-4787-4760} \and
Tae Hyung Kim\inst{5}\orcidID{0000-0001-5881-7265} \and
Berkin Bilgic\inst{3,4}\orcidID{0000-0002-9080-7865} \and
Jaejin Cho\inst{1}$^*$\orcidID{0000-0001-5672-6765}}
\authorrunning{N. Han et al.}
%
\institute{Department of Artificial Intelligence and Robotics, Sejong University, Seoul 05006, South Korea \and
Institute of Neuroscience and Medicine 4, INM-4, Forschungszentrum Jülich, Jülich, Germany \and
Athinoula A. Martinos Center for Biomedical Imaging, Charlestown, MA 02129, USA \and
Harvard Medical School, Boston, MA 02115, USA \and
Department of Computer Engineering, Hongik University, Seoul 04066, South Korea\\
\email{jaejincho@sejong.ac.kr}}

  
\maketitle              

\def\thefootnote{$\dagger$}\footnotetext{These authors contributed equally to this work.}
\def\thefootnote{$*$}\footnotetext{Corresponding author: jaejincho@sejong.ac.kr}
\def\thefootnote{\arabic{footnote}} 

\begin{center}
\textit{Preprint. Under review}
\end{center}

\begin{abstract}
Echo-planar imaging (EPI) remains the cornerstone of diffusion MRI, but it is prone to severe geometric distortions due to its rapid sampling scheme that renders the sequence highly sensitive to $B_{0}$ field inhomogeneities. While deep learning has helped improve MRI reconstruction, integrating robust geometric distortion correction into a self-supervised framework remains an unmet need. To address this, we present FINDER (Field-Integrated Network for Distortion-free EPI Reconstruction), a novel zero-shot, scan-specific framework that reformulates reconstruction as a joint optimization of the underlying image and the $B_{0}$ field map. Specifically, we employ a physics-guided unrolled network that integrates dual-domain denoisers and virtual coil extensions to enforce robust data consistency. This is coupled with an Implicit Neural Representation (INR) conditioned on spatial coordinates and latent image features to model the off-resonance field as a continuous, differentiable function. Employing an alternating minimization strategy, FINDER synergistically updates the reconstruction network and the field map, effectively disentangling susceptibility-induced geometric distortions from anatomical structures. Experimental results demonstrate that FINDER achieves superior geometric fidelity and image quality compared to state-of-the-art baselines, offering a robust solution for high-quality diffusion imaging.

\end{abstract}
\section{Introduction}

Diffusion MRI (dMRI) is essential for non-invasively mapping tissue microstructure \cite{le2015diffusion}. To accommodate the lengthy acquisitions required for multi-directional diffusion encoding, Echo-Planar Imaging (EPI) is universally adopted for its high acquisition speed. However, the long readout train of EPI results in a limited bandwidth along the phase-encoding direction, making the acquisition highly susceptible to $B_0$ field inhomogeneities. This leads to severe geometric distortions and signal intensity modulations, which are further exacerbated in high-resolution imaging due to increased phase error accumulation.

To mitigate geometric distortions without explicit $B_0$ mapping, reversed phase-encoding (blip-up/down) strategies using FSL-TOPUP \cite{andersson2003correct,smith2004advances} are conventionally used. The Blip-Up and -Down Acquisition (BUDA) framework \cite{liao2019highly,zhang2022blip,liao2023high} represents the state-of-the-art by combining TOPUP field estimation and parallel imaging with structured low-rank constraints (e.g., MUSSELS \cite{Mani2017-bo} or S-LORAKS \cite{Haldar2014-ov,Haldar2016-ou,kim2018loraks,lobos2021robust}) to jointly reconstruct EPI data. Despite its robust performance, this pipeline is inherently limited by the initial SENSE \cite{Pruessmann1999-jb} image quality used in the TOPUP field estimation. At high acceleration factors, residual aliasing artifacts inevitably propagate into the field estimation, ultimately degrading the final reconstruction fidelity. 

Deep learning (DL) effectively mitigates the computational bottlenecks of iterative methods \cite{hammernik2018learning,eo2018kiki,Akcakaya2019-lu,han2019k}. Specifically, unrolled architectures like Model-Based Deep Learning (MoDL) \cite{Aggarwal2019-er,Aggarwal2020-wy} enforce strict data consistency (DC), providing a rapid, effective alternative to structured low-rank methods for multi-shot EPI. Despite this superior performance, conventional DL relies heavily on large-scale, artifact-free ground-truth datasets, which are difficult to acquire in dMRI. To circumvent this, Zero-Shot Self-Supervised Learning (ZS-SSL) \cite{Yaman2020-ei,Yaman2022-it} offers a compelling paradigm, enabling scan-specific network optimization without external training data.

Despite these advances, extending DL to geometric distortion correction remains challenging. A recent work, the Forward-Distortion Network (FD-Net) \cite{zaid2024fd}, leverages physics-driven constraints to jointly estimate the field map and underlying image. However, as a magnitude-domain post-processing step, its efficacy is highly dependent on the initial blip-up/down parallel imaging quality \cite{Pruessmann1999-jb,Griswold2002-cs}. Similarly, the Physics-informed Implicit Neural Representation (PINR) \cite{huang2025physics} jointly estimates $B_0$ field map and images with rotating views. While PINR shows high-fidelity results on simulated data, translating it to \textit{in vivo} dMRI is fundamentally limited by unaddressed dynamic shot-to-shot phase variations and eddy currents.

To address these challenges, we propose the Field-Integrated Network for Distortion-free EPI Reconstruction (FINDER), a novel zero-shot framework. FINDER synergistically couples a physics-guided unrolled network for image recovery with a coordinate-based INR for improved field estimation. Unlike discrete CNNs, INRs map spatial coordinates directly to signal values, acting as continuous, differentiable functions. This inherent geometric smoothness makes them exceptionally well-suited for modeling $B_0$ field inhomogeneities, which vary smoothly across space.

Our main contributions are: (1) We propose FINDER, a scan-specific, zero-shot framework that jointly reconstructs highly accelerated blip-up/down EPI data while simultaneously correcting geometric distortions without external datasets. (2) We introduce a robust field update strategy, initializing from TOPUP and refining via an INR to accurately capture field inhomogeneities. (3) To ensure high-fidelity image recovery, we employ a physics-guided unrolled network that integrates dual-domain denoisers and virtual coil extensions \cite{blaimer2009virtual}. We validate FINDER through \textit{in vivo} experiments using 5-fold accelerated dMRI data.

\section{Method}

\subsection{Problem Formulation of Joint Reconstruction}

To mathematically formalize geometric distortion correction in dMRI, the acquired k-space data $y_{d,m}$ for the $m^{th}$ shot in the $d^{th}$ direction is modeled as $y_{d,m} = \mathcal{F}_{m}\mathbf{E}_{d}\mathbf{C}x_{d}$, where $x_{d}$ is the distortion-free image, $\mathbf{C}$ is the coil sensitivity map, $\mathbf{E}_{d}$ is the off-resonance phase operator, and $\mathcal{F}_{m}$ is the undersampled Fourier transform. Conventional methods treat $\mathbf{E}_{d}$ as a fixed prior, which inevitably propagates initial estimation errors into the final image.

To address this, FINDER reformulates the reconstruction as an optimization problem utilizing learned components for both the field map and the image prior. At inference, the final distortion-free image $\hat{x}_{d}$ is obtained by solving:

\begin{equation}
{\hat{x}_{d},G_{\hat{\theta}}} = \underset{x_d,G_{\hat{\theta}}}{\mathrm{argmin}}  \;  \sum_{m=1}^{M} \|\mathcal{F}_{m} \mathbf{E}_{d} (G_{\hat{\theta}}) \mathbf{C} x_{d} - y_{d,m}\|_{2}^{2} + \lambda \mathcal{R}_{\text{DL}}(x_{d}; \hat{\phi})
\end{equation}

\noindent where $G_{\hat{\theta}}$ is the optimized INR network modeling the field map, and $\mathcal{R}_{\text{DL}}(x_d; \hat{\phi})$ imposes the learned regularization prior parameterized by the optimized unrolled network weights $\hat{\phi}$.

\subsection{Network Architecture}

\begin{figure}[!htbp]
\includegraphics[width=\textwidth]{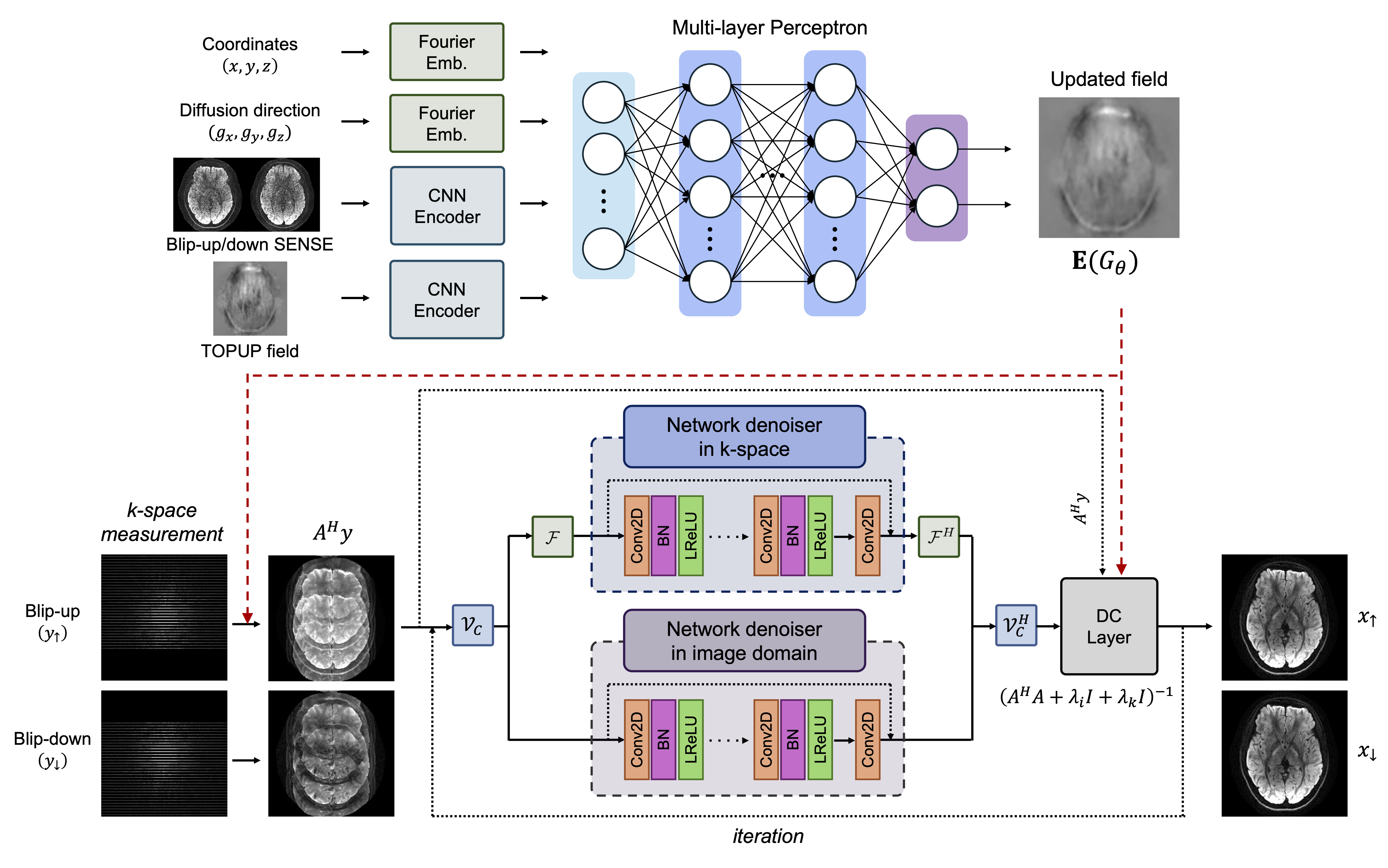}
\caption{Overview of the proposed FINDER framework. The upper part presents the INR network for field map updates, while the lower part details the physics-guided unrolled image reconstruction network.} \label{fig1}
\end{figure}

As illustrated in Fig.~\ref{fig1}, FINDER integrates two specialized neural modules: an INR-based field estimator and a deep unrolled image reconstructor.

\subsubsection{Physics-Informed INR Field Estimator.} 

To capture the continuous and smooth nature of the $B_0$ field map, we employ a coordinate-based Multi-layer Perceptron (MLP). The inputs comprise: (1) spatial coordinates $(x, y, z)$, (2) the diffusion gradient vector $(g_x,g_y,g_z)$, and (3) latent feature maps extracted from the preliminary blip-up/down SENSE images and the initial TOPUP field via a lightweight CNN encoder. To better represent high-frequency details, we apply multi-scale Fourier feature embeddings \cite{tancik2020fourier,mildenhall2021nerf} to the coordinate inputs. The concatenated features are fed into a 5-layer MLP to directly predict the field map specific to each diffusion direction.

\subsubsection{Dual-Domain Unrolled Reconstruction Network.}

While the INR corrects geometric distortions, the reconstructed image quality is governed by the regularization prior. To overcome the limitations of hand-crafted priors, we propose a deep unrolled network incorporating a Virtual Coil ($\mathcal{V}_C$) strategy \cite{blaimer2009virtual} and dual-domain regularization. Unlike standard single-domain methods, our regularizer simultaneously targets artifacts across two domains: an image-domain U-Net ($\mathcal{N}_{\text{i}}$) removes spatially varying noise, while a k-space-domain network ($\mathcal{N}_{\text{k}}$) restores missing frequency components. Specifically, the regularization term is formulated as:

\begin{equation}
\mathcal{R}_{\text{DL}}(x_d) = \lambda_{\text{i}} \left\| x_d - \mathcal{V}_C^H \mathcal{N}_{\text{i}} (\mathcal{V}_C x_d) \right\|_2^2 + \lambda_{\text{k}} \left\| x_d - \mathcal{F}^{H} \mathcal{V}_C^H \mathcal{N}_{\text{k}} (\mathcal{V}_C \mathcal{F} x_d) \right\|_2^2
\end{equation}

\noindent where $\mathcal{N}_{\text{i}}$ and $\mathcal{N}_{\text{k}}$ denote the CNN-based residual networks operating in the image and k-space domains, respectively. These components are integrated into a model-based unrolled architecture inspired by MoDL \cite{Aggarwal2019-er,Aggarwal2020-wy}. Instead of explicitly solving complex regularized objective functions, the optimization in each unrolled iteration alternates between a CNN-based dual-domain denoising step and a conjugate gradient-based DC step. The DC step efficiently enforces strict adherence to the physics of the acquisition model by dynamically incorporating the estimated field map into the forward model operator $\mathbf{A} = \mathcal{F} \mathbf{E}(G_{\hat{\theta}}) \mathbf{C}$.

\subsection{Zero-Shot Optimization Strategy}

As acquiring high-quality, distortion-free ground-truth data is inherently challenging and time-consuming in clinical dMRI settings, FINDER is optimized in a fully zero-shot, self-supervised manner using only acquired k-space data. We adopt the ZS-SSL strategy \cite{Yaman2020-ei,Yaman2022-it}, partitioning k-space into distinct masks for training input, loss calculation, and validation. 

To resolve the highly ill-posed nature of joint optimization, we employ a structured alternating minimization strategy. First, the INR ($\theta$) and spatial encoders are trained to learn the initial TOPUP field map. Second, with $\theta$ frozen, the reconstruction network ($\phi$) is updated for a single epoch by minimizing the standard ZS-SSL $L_1$ and $L_2$ DC losses on the unseen loss mask, focusing on initial aliasing artifact removal. Finally, we alternately update $\theta$ and $\phi$ while freezing the other. To refine the field estimation ($\theta$), we minimize a combination of $L_2$ intensity and $L_1$ gradient differences between the corrected blip-up and blip-down images. For the reconstruction update ($\phi$), the objective minimizes the ZS-SSL DC loss on the loss mask alongside the $L_2$ difference between blip-up/down reconstructions.

\section{Experiments and Results}
\subsection{Experimental Setup}

\textit{In vivo} dMRI data were acquired on a 3T Siemens Prisma scanner equipped with a 32-channel head coil. We acquired 32 diffusion-encoded directions using a 2-shot EPI sequence. Each shot was accelerated by a factor of 5 ($R=5$) and employed 75\% Partial Fourier. The imaging parameters were: FOV = $220 \times 220 \times 128 \text{ mm}^3$, spatial resolution = $1 \times 1 \times 4 \text{ mm}^3$, TR = 3.5 s, and TE = 59 ms. To establish a pseudo-ground truth (GT), we additionally acquired 5-shot blip-up and 5-shot blip-down EPI data, reconstructed using BUDA-LORAKS.

FINDER was implemented in PyTorch and optimized on an NVIDIA H100 GPU. The dual-domain denoising networks employ a 16-layer CNN architecture with $3 \times 3$ filters and 46 feature channels. The unrolled architecture consists of 8 iterations, with the DC block utilizing 10 conjugate gradient steps. The INR network consists of 5-layer MLP with 512 feature channels. For zero-shot optimization, a single unified network was trained across all 32 diffusion directions for 3 epochs using the Adam optimizer with a learning rate of 1e-5. We generated 10 training and 1 validation masks per direction. Exemplar data and code are available at: \url{https://github.com/jaejin-cho/FINDER-MRI}.

We compared FINDER against established state-of-the-art methods: FSL-TOPUP, BUDA-LORAKS, and FD-Net. For FSL-TOPUP and FD-Net, field maps were estimated from the initial SENSE reconstructions of blip-up/down images. For the baseline BUDA-LORAKS, the TOPUP-estimated field map was used to reconstruct the final distortion-corrected images from the 5-fold accelerated data. For FD-Net, although the original FD-Net utilizes weights pre-trained on external datasets \cite{van2013wu}, we trained the network from scratch in a scan-specific manner to evaluate its performance at the native resolution. Across all methods, multi-shot images were merged using a real-value combination \cite{eichner2015real}, and downstream diffusion analysis was performed using the FSL toolbox \cite{andersson2003correct,smith2004advances,jenkinson2012fsl}.

\begin{figure}[!htbp]
\includegraphics[width=\textwidth]{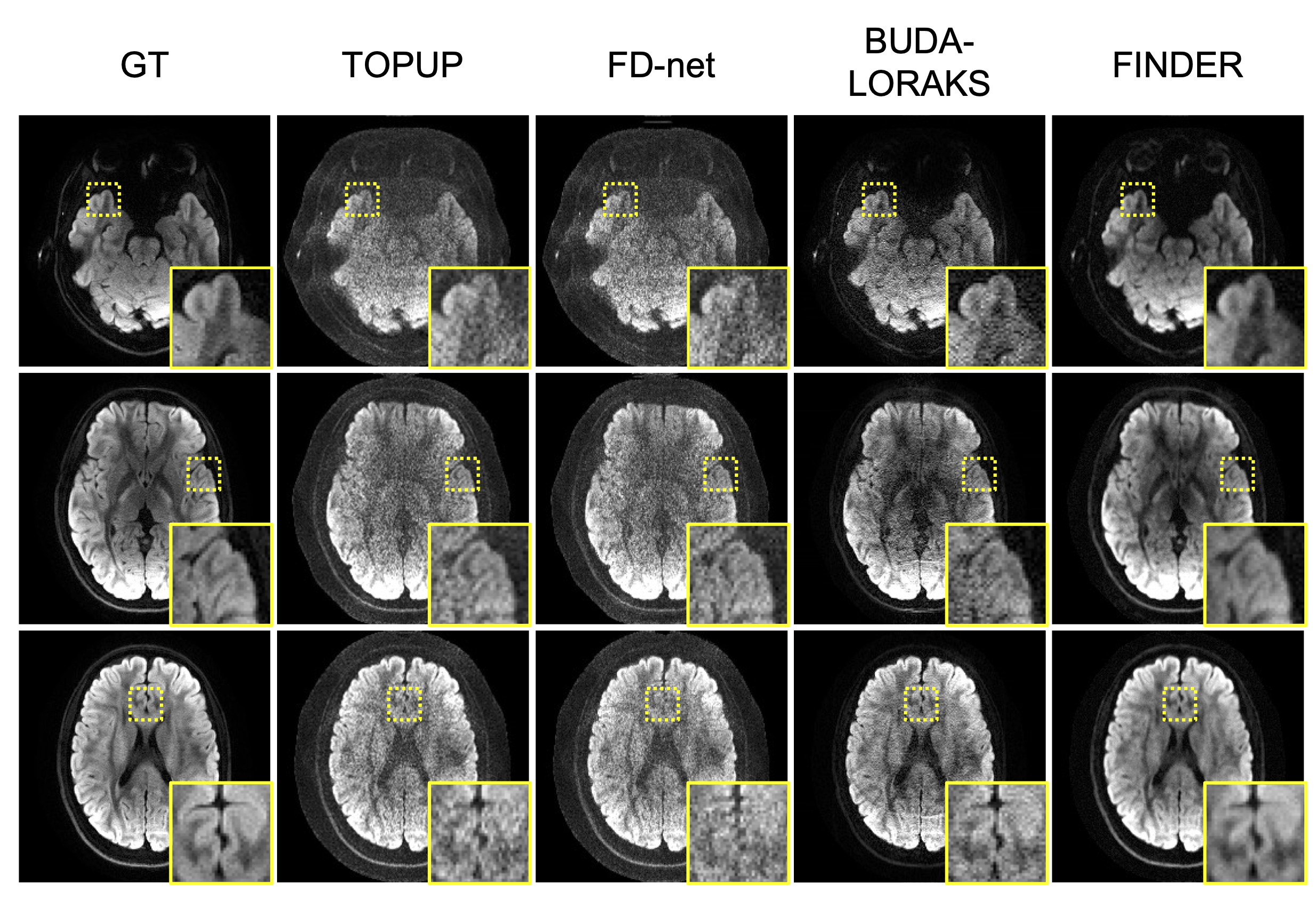}
\caption{Visual comparison of the diffusion-weighted images at 5-fold acceleration (single direction). Yellow insets highlight baseline artifacts effectively suppressed by FINDER. } \label{fig2}
\end{figure}

\subsection{High-Fidelity Distortion-Free Reconstruction}

Fig.~\ref{fig2} presents a visual comparison of the reconstructed slices for 5-fold in-plane accelerated \textit{in vivo} data. At this high acceleration factor, because TOPUP and FD-net are purely post-processing methods, their performance is inherently limited by the initial magnitude image quality derived from SENSE reconstruction, leading to conspicuous residual folding artifacts and noise amplification. While BUDA-LORAKS attempts to mitigate noise amplification, it still suffers from residual folding artifacts. In contrast, FINDER effectively addresses both noise and folding artifacts, successfully restoring anatomically faithful geometry and microstructural details. FINDER yields reconstructions highly consistent with the GT, effectively suppressing both aliasing and geometric distortions, which is further supported by the quantitative evaluation in Table ~\ref{tab1}. FINDER consistently outperforms all baseline methods across all metrics, including Normalized Root Mean Squared Error (NRMSE), Local Cross-Correlation (LCC), Structural Similarity Index Measure (SSIM) \cite{wang2004image}, High-Frequency Error Norm (HFEN), and Learned Perceptual Image Patch Similarity (LPIPS) \cite{zhang2018unreasonable}.

\begin{table}[!htbp]
\centering
\caption{Quantitative metrics averaged across all slices and diffusion directions}\label{tab1}
\begin{tabularx}{\textwidth}{lYYYY} 
\toprule
Metric                  & TOPUP     & FD-net    & BUDA-LORAKS   & FINDER \\
\midrule
NRMSE $\downarrow$      & 0.3426    & 0.3471    & 0.2066        & \textbf{0.1311}    \\
LCC $\uparrow$          & 0.9520    & 0.9410    & 0.9778        & \textbf{0.9904}   \\
SSIM $\uparrow$         & 0.6037    & 0.5579    & 0.7949        & \textbf{0.8451}   \\
HFEN $\downarrow$       & 0.5334    & 0.6305    & 0.3702        & \textbf{0.2669}   \\
LPIPS $\downarrow$      & 0.2934    & 0.3485    & 0.1646        & \textbf{0.0925}   \\
\bottomrule
\end{tabularx}
\end{table}

\begin{figure}[!htbp] %
\includegraphics[width=\textwidth]{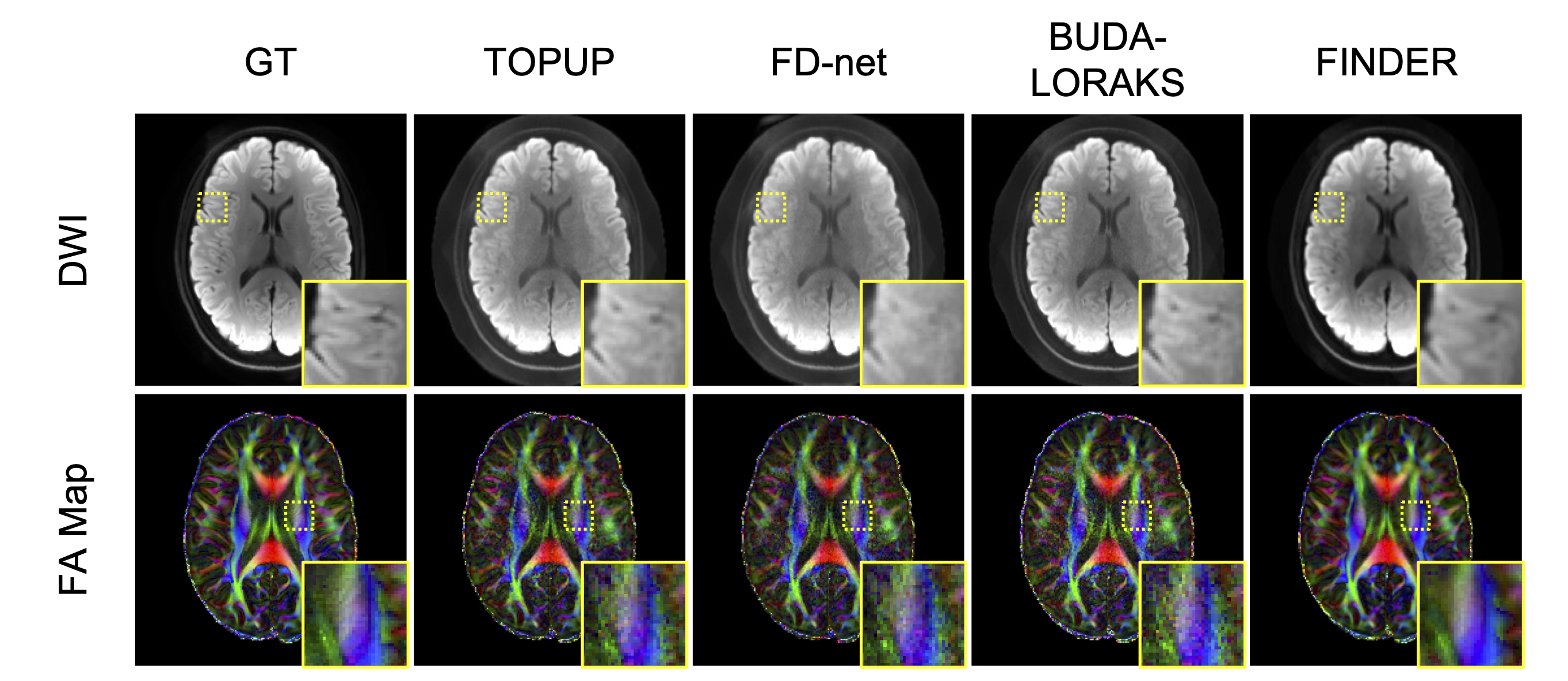}
\caption{Visual comparison of the average DWIs across 32 directions and the corresponding FA maps. Yellow insets highlight severe noise amplification and artifacts in baseline methods, which are effectively suppressed by FINDER.} 
\label{fig3}
\end{figure}

\subsection{Diffusion Analysis and Field Refinement}

Fig.~\ref{fig3} presents the average diffusion-weighted images (DWIs) and corresponding fractional anisotropy (FA) maps derived from 32 directions. While BUDA-LORAKS and FINDER produce high-fidelity average DWIs, TOPUP and FD-net suffer from severe noise amplification. Notably, in the FA maps, all baseline methods exhibit conspicuous noise in the central brain regions. In contrast, FINDER effectively mitigates these artifacts, yielding robust and reliable microstructural estimations highly consistent with the ground truth.

\begin{figure}[!htbp]
\includegraphics[width=\textwidth]{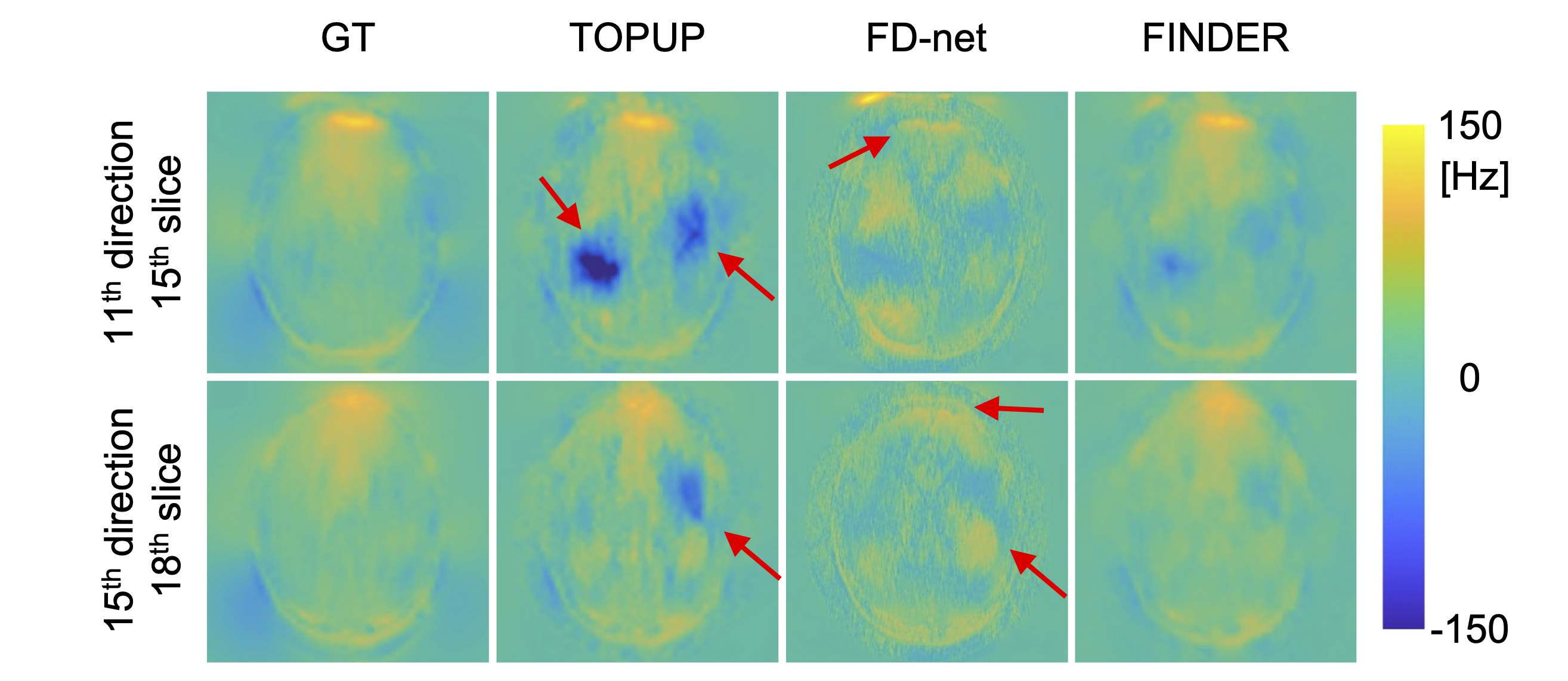}
\caption{Visual comparison of the estimated field maps. TOPUP and FD-net estimate the field maps based on the highly aliased initial SENSE reconstructions. In contrast, FINDER dynamically refines the initial TOPUP field map using the proposed INR network. Red arrows indicate severe estimation errors, which are effectively mitigated by FINDER. The RMSEs of field maps in the brain region are 19.43 Hz, 31.07 Hz, and 13.57 Hz for TOPUP, FD-net, and FINDER, respectively.} 
\label{fig4}
\end{figure}

To validate the accuracy of the field estimation, Fig.~\ref{fig4} compares the refined field maps. TOPUP exhibits significant estimation errors induced by the 5-fold acceleration. FD-net faced challenges in yielding an accurate field map under solely scan-specific, self-supervised training. FINDER effectively mitigates these errors by dynamically refining the initial TOPUP field via the proposed INR network. The continuous and differentiable nature of the INR allows FINDER to capture spatially smooth field inhomogeneities more accurately, leading to superior geometric correction.

\section{Discussion and Conclusion}

In this study, we proposed FINDER, a novel framework for joint image reconstruction and INR-based field refinement in highly accelerated MRI. Our \textit{in vivo} results demonstrate that FINDER effectively mitigates severe geometric distortions and aliasing artifacts without relying on high-quality initial images. By successfully restoring anatomically faithful geometry, FINDER significantly outperforms conventional post-processing state-of-the-art methods in both qualitative and quantitative evaluations.

Despite its superior reconstruction fidelity, several limitations remain to be addressed. First, while FINDER strictly adheres to the physical forward model of image distortion, the inherent line-by-line phase modulation during the Fourier transform increases computational complexity. This results in substantial VRAM requirements and intensive computing power, leading to a prolonged training time of approximately two days per epoch to process the entire 3D volume across all 32 diffusion directions. At inference, reconstructing full volumes in all diffusion directions takes approximately 1 hour on the NVIDIA H100 GPU, while BUDA-LORAKS took approximately 4.5 hours on AMD Ryzen 9 9950X CPU. Such computational bottlenecks will be mitigated by adopting meta-learning strategies or advanced parallel computing techniques. Second, the joint optimization of two intricately linked networks, the unrolled reconstructor and the INR field estimator, presents numerical challenges. Even with an alternating minimization strategy, the convergence remains sensitive to hyperparameter tuning, including loss balancing and learning rate scheduling.

Future work will focus on reducing computational overhead by utilizing image-domain displacement estimation combined with Jacobian modulation. This approach, mirroring strategies in TOPUP and FD-net, promises significantly faster implementation while preserving physics-based accuracy. In conclusion, FINDER provides a highly robust, distortion-free reconstruction solution, paving the way for more reliable and highly accelerated dMRI in clinical practice.

\begin{credits}

\subsubsection{\ackname} 
This work was supported by the National Research Foundation of Korea (NRF) grant funded by the Korea government (MSIT) (No. RS-2025-00555277, No. RS-2025-24534878).

\subsubsection{\discintname}
The authors have no competing interests to declare that are relevant to the content of this article.


\end{credits}

%
%
%

\bibliographystyle{splncs04}
\bibliography{mybibliography}




\end{document}